# Towards targeted kinetic trapping of organic-inorganic interfaces: A computational case study

Anna Werkovits, Andreas Jeindl, Lukas Hörmann, Johannes J. Cartus, Oliver T. Hofmann*

Institute of Solid State Physics, TU Graz, NAWI Graz, Petersgasse 16/II, 8010 Graz, Austria

**ABSTRACT:** Properties of inorganic-organic interfaces, such as their interface dipole, strongly depend on the structural arrangements of the organic molecules. A prime example is tetracyanoethylene (TCNE) on Cu(111), which shows two different phases with significantly different work functions. However, the thermodynamically preferred phase is not always the one that is best suited for a given application. Rather, it may be desirable to selectively grow a kinetically trapped structure. In this work, we employ density functional theory and transition state theory to discuss under which conditions such a kinetic trapping might be possible for the model system of TCNE on Cu. Specifically, we want to trap the molecules in the first layer in a flat-lying orientation. This requires temperatures that are sufficiently low to suppress the re-orientation of the molecules, which is thermodynamically more favorable for high dosages, but still high enough to enable ordered growth through diffusion of molecules. Based on the temperature-dependent diffusion and re-orientation rates, we propose a temperature range at which the re-orientation can be successfully suppressed.

## 1. INTRODUCTION

Metal-organic interfaces act as basis for a variety of possible nanotechnological applications, such as molecular switches,[1,2] thermoelectrics,[3,4] memories,[5] transistors,[6–8] or spintronic devices.[9] Due to the advances in computational material design, possibilities for developing functional interfaces with tailored physical properties and functionalities have increased in the last decades.[10,11] However, the functionality of these interfaces does not depend on the choice of the metal and the organic component alone. Rather, also the structure the organic component assumes on the surface plays a decisive role. A prime example are molecular acceptors that undergo a (coverage-dependent) re-orientation from flat-lying to upright-standing positions, like hexaazatriphenylene-hexacarbonitrile (HATCN) and dinitropyrene-tetraone (NO2-Pyt) on Ag(111).[12,13] Because the electron affinity of organic films depends on their orientation,[14] this is accompanied by significant changes of the charge transfer and interface work functions.[13,15] In the two examples above, the structural transition causes a change of the work function of more than 1 eV, illustrating how important control over the structure is.

In principle, such control can be achieved by identifying process conditions that allow to grow the target structure in thermodynamic equilibrium.[16,17] In practice, however, often kinetically trapped phases appear, especially when preparing interfaces using physical vapor deposition.[18] This is because kinetics plays a major role: Following Ostwald's rule of stages,[19] thermodynamically less stable structures form first. Whether the transition to a more stable structure occurs or whether the structure becomes kinetically trapped depends on the energetic barriers and the corresponding transition rates. Therefore, we can make a virtue out of a necessity by explicitly utilizing kinetic trapping to grow structures out of thermodynamical equilibrium: In theory, controlled formation of a kinetically trapped structure should be possible by selecting a deposition temperature where the rate for the phase transition to a thermodynamically more stable structure is slower than the speed at which the trapped structure grows. This requires profound knowledge of a) the underlying transition mechanisms, and b) the ability of the molecules to diffuse and aggregate, i.e. to form a seed for a different structure vis-à-vis to continue growing in the less thermodynamically stable form.

In this work, we perform a first step to predict controlled growth of the model system tetracyanoethylene (TCNE) on Cu(111). While being computationally more tractable than its cousins HATCN and $NO_2$-PyT on Ag(111), it reveals an even larger change in work function. When increasing the dosage of TCNE, the system undergoes a re-orientation from flat-lying to upright-standing molecules,[20] which leads to a work function increase of approximately 3 eV. When continuing growth, a second layer of TCNE forms on top of the first, standing, monolayer.[20]

As the layer in direct contact with the surface is the decisive factor for the properties of the interface,[7] it is highly interesting to study how the re-orientation within the first layer could be suppressed for high dosages. To take a first step in predicting how the re-orientation of TCNE on Cu(111) could be prevented, here we study, by first principles, under which conditions the re-orientation can be kinetically suppressed altogether already for individual TCNE molecules, i.e. when not even a single molecule is able to adopt the upright-standing geometry within a reasonable timescale. However, computing TCNE on Cu(111) faces a fundamental challenge: The re-orientation on the surface substantially alters the way the molecules interact with the surface.[20] This includes charge transfer and the connected re-hybridization of molecular and metal orbitals. These orbital re-hybridizations are not covered by state-of-the-art force field-approaches, rendering them (and, by extension, molecular dynamics simulations) inapplicable here. Instead, we use dispersion-corrected density functional theory (for details see Methods), to obtain minimum energy paths and transition states by the nudged elastic band method.[21,22] This method was previously successfully employed to study diffusion processes of inorganic-organic interfaces.[23,24] Applying harmonic transition state theory,[25,26] we can further determine temperature-dependent rates of diffusion and re-orientation. This allows us to estimate a temperature range at which the re-orientation is suppressed while further growth of lying seeds is still supported - resulting in a kinetic trapping of lying TCNE.

## 2. RESULTS AND DISCUSSION

### 2.1 Transition paths and barriers

Arguably, the ability of the system to undergo a phase transition depends on the molecules' ability to diffuse on the surface and, more importantly, on the rate at which they can change their orientation. Generally, re-orientation processes happen spontaneously and are typically energetically driven: Once a sufficient number of molecules aggregate in the upright-standing geometry (i.e., the exceed a so-called critical nucleus size), this geometry becomes energetically favorable compared to flat lying geometries. The number of molecules for this critical cluster size varies from system to system (and it thought to be between 3-10).[27–30] However (collaborative re-orientations notwithstanding), it is clear that at least a single molecule must re-orient (i.e., some molecule has to make the first step). This provides a limit to the rate at which critical clusters can form in the first place.

Consequently, a useful first step is to investigate these processes for individual molecules, rather than directly studying transitions between full close-packed structures. In our case, this is justified because both, the most favorable flat-lying and the most favorable upright-standing structure, consist of molecular geometries that would also be stable local minima on their own due to the strong molecule-substrate interactions. In addition, this reduced complexity enables studying kinetic processes at a feasible cost. Therefore, we omit multi-molecule processes that include intermolecular interactions, such as the initial nucleation, attachment and detachment processes from an island, and Ehrlich-Schwoebl barriers. Instead, we focus exclusively on two fundamental aspects in the low coverage growth regime: The diffusion and the re-orientation of individual molecules on the surface.

Before we explain which transitions we compute in detail, we briefly introduce the stable adsorption geometries of individual TCNE molecules on Cu(111) and the two structures we are interested in. This information was previously provided by *Egger, et al.*[20] and is repeated here, as the local geometries are the starting points for all our further computations.

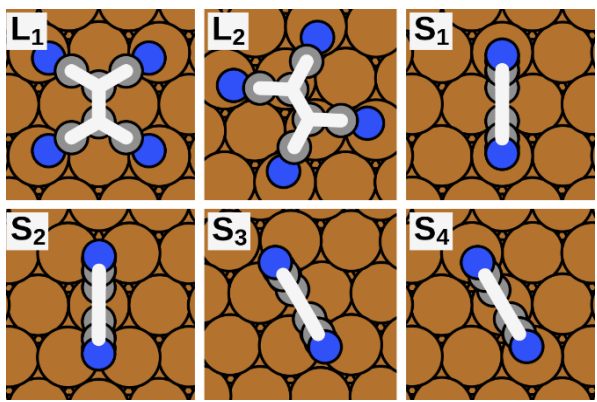

**Figure 1.** Top view of stable adsorption geometries for TCNE on Cu(111).[20] The orange spheres represent the Cu atoms of the substrate, whereas the grey and blue spheres are the C and N atoms, respectively. The white overlay is a reduced representation used in further plots. In addition, the directions of the Cu(111) substrate are provided.

The most favorable way for individual TCNE molecules to adsorb on the Cu(111) surface is in a "flat-lying" position. For this case, there are two different possible adsorption geometries (i.e., local minima) for the molecule, which we will further denote as $L_1$ and $L_2$ (see Figure 1). There are also four "upright-standing" adsorption geometries, denoted as $S_1$ to $S_4$ (also shown in Figure 1). Their energies are more than 0.5 eV higher, i.e. less stable.

In a full monolayer, at low coverages the energetically most favorable structure consists exclusively of flat-lying molecules in the $L_1$ position, while its pendant for high coverages includes the upright-standing positions $S_1$, $S_3$ and $S_4$ (for details see Supporting Information).[20] It is clear that for the flat-lying structure to grow, lying molecules must be able to diffuse on the surface ("lying diffusion"). The formation of the upright standing structure requires that molecules can re-orient from lying to standing ("re-orientation"), and, potentially, that the standing molecules can also diffuse on the surface ("standing diffusion").

To study diffusion and re-orientations, we create a representative set of distinct transitions between pairs of adsorption geometries (including their rotational and translational symmetry equivalents). When naïvely accounting for, e.g., three rotational and three translational symmetry equivalents for each adsorption geometry, we would already get $\binom{36}{2} = 630$ transitions. Nevertheless, these will decompose into a manageable set of symmetry equivalent "elementary" transitions, i.e., transitions that possess exactly one transition state and therefore proceed in a single step. By knowing these elementary transitions, pathways can be constructed by linking individual elementary paths in a way that yields the lowest energetic barrier for the total transition. To efficiently obtain the most relevant elementary transitions, sets of start- and endpoints are selected based on two concepts: Firstly, we restrict the selection to adsorption geometries in adjacent adsorption sites (i.e. translationally equivalent adsorption positions which are anchored on neighboring Cu atoms on the surface), implying that the adsorbate centers are at maximum one Cu-lattice constant apart. In other words, we neglect so called "long jumps". This is warranted because there is evidence that such long jumps are improbable for moderate to low temperatures and for small molecules.[31,32] Secondly, we assume that for moderate to low temperatures kinetics is mainly dominated by transitions including the adsorption geometry with the lowest energy in its class either as start and/or end point (i.e. geometry $L_1$ for flat-lying adsorbates and $S_1$ for the upright-standing ones). This is warranted because low-energy structures also tend to have low-energy barriers due to their wide basin of attraction.[33–36] This assumption is also confirmed in hindsight by our results (*vide infra*). Therefore, we initialized ten transitions as depicted in Figure 2 (a-j). Although this does not provide all possible transitions, with this strategy we expect to obtain the most dominant and thus limiting processes of the distinct transition regimes. To conveniently indicate transitions from an adsorption geometry **A** to another adsorption geometry **B**, we use a notation of the form **A→B**. Hereby, **A→B** is simply referred to as "forward" transition, while the transition with inverted initial and final states (**B→A**) is denoted as "reverse" transition.

To model the diffusion of lying TCNE molecules, we consider four possible transitions: Three different transitions that go directly from one $L_1$ to another $L_1$ at a different adsorption site (Figure 2a-c), and the transition from $L_1$ to the nearest $L_2$ geometry (Figure 2d). The different $L_1$→$L_1$ transitions consist of two direct transitions to neighboring adsorption sites (shown in Figure 2a and b) and one transition to a symmetry-equivalent rotated geometry (Figure 2c).

For the re-orientation, we consider transitions from the lying minimum $L_1$ to the standing end points $S_1$ (Figure 2e), $S_3$ (Figure 2f) and $S_4$ (Figure 2g), i.e. to each of the geometries contained in the upright-standing structure.

For the diffusion of standing molecules, the most favorable adsorption geometry of this class, $S_1$, is always chosen as initial minimum that transitions into either $S_1$ (Figure 2l), $S_2$ (Figure 2i), $S_3$ (Figure 2j) or $S_4$ (Figure 2k).

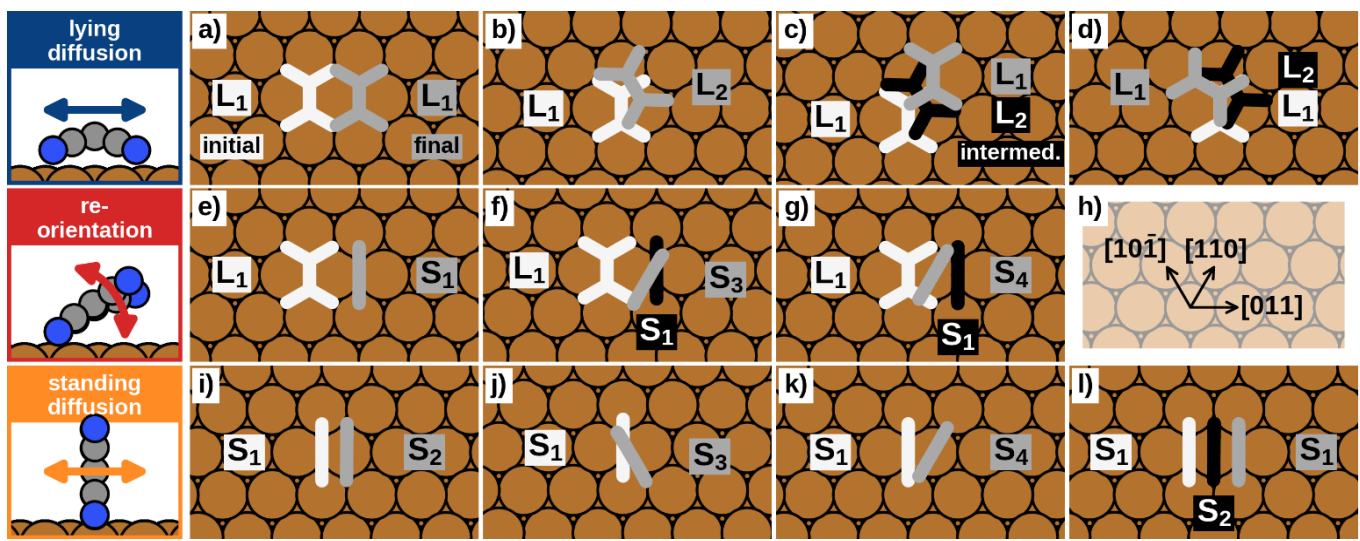

**Figure 2.** Selected start- (white) and endpoints (grey) of the transitions for the lying diffusion (a-d), re-orientation (e-g) and the standing diffusion (i-l). Intermediate steps of multistep transitions (non-elementary transitions) are colored black. For a clear representation, the molecule geometries are displayed in a reduced form that omits the nitrogen atoms, corresponding to the overlay in Figure 1. In (h), directions of the substrate lattice are stated.

In the course of our computations, we found that five of the eleven initialized transitions occur as multi-step processes, i.e. they include another adsorption geometry and are, therefore, a combination other transition processes: Two of the three **L₁→L₁** diffusion transitions (Figure 2b, c) proceed *via* the adsorption geometry **L₂** and thus reduce to consecutive transitions of **L₁→L₂** and **L₂→L₁** (Figure 2d). Therefore, the remaining **L₁→L₁** transition (Figure 2a) uniquely denotes the direct transition. In addition, the re-orientations **L₁→S₃** (Figure 2f) and **L₁→S₄** (Figure 2g) proceed *via* **S₁**, inferring that the main re-orientation process is **L₁→S₁** (Figure 2e). By investigating **L₁→S₁** in more detail (discussed later), we also found that this transition proceeds *via* a hitherto overlooked intermediate minimum (**M**). Upright-standing diffusions of **S₁** to all symmetry equivalent **S₁** in adjacent adsorption sites (e.g. along the directions <011>) occur as multistep processes over **S₂**, **S₃** and/or **S₄** (see Supporting Information). All remaining transitions possess exactly one transition state, i.e. occur as "elementary" transitions. In total, the transition states and minimum energy paths of seven elementary transitions were obtained: Lying TCNE molecules can either diffuse along <011> directions (**L₁→L₁**) or perform rotations (**L₁→L₂**). The observed process of re-orientation from a flat-lying to an upright-standing position occurs consecutively *via* **L₁→M** and **M→S₁**, as discussed later in more detail. For the standing diffusion, the motion in straight lines perpendicular to the molecular plane (similar to a walking motion) is enabled *via* **S₁→S₂**, while rotation between the directions <011>, <110> and <10$\bar{1}$> of the Cu(111) surface occurs by **S₁→S₃** and **S₁→S₄**. For more insight, animated GIFs are provided in the Supporting Information. Figure 3 shows an overview containing the main geometric characteristics. This includes the initial and final adsorption geometries, as well as the positions and the explicit geometries of the obtained transition states. The only exception is the transition **L₁→S₁** (Figure 3c), where the intermediate minimum **M** is provided instead. The corresponding energy barriers, as well as the absolute adsorption energies of the initial states, the transition states and the final states are summarized in Table 1 and visualized in Figure 4. The detailed energy paths of all transitions are visualized in the Supporting Information.

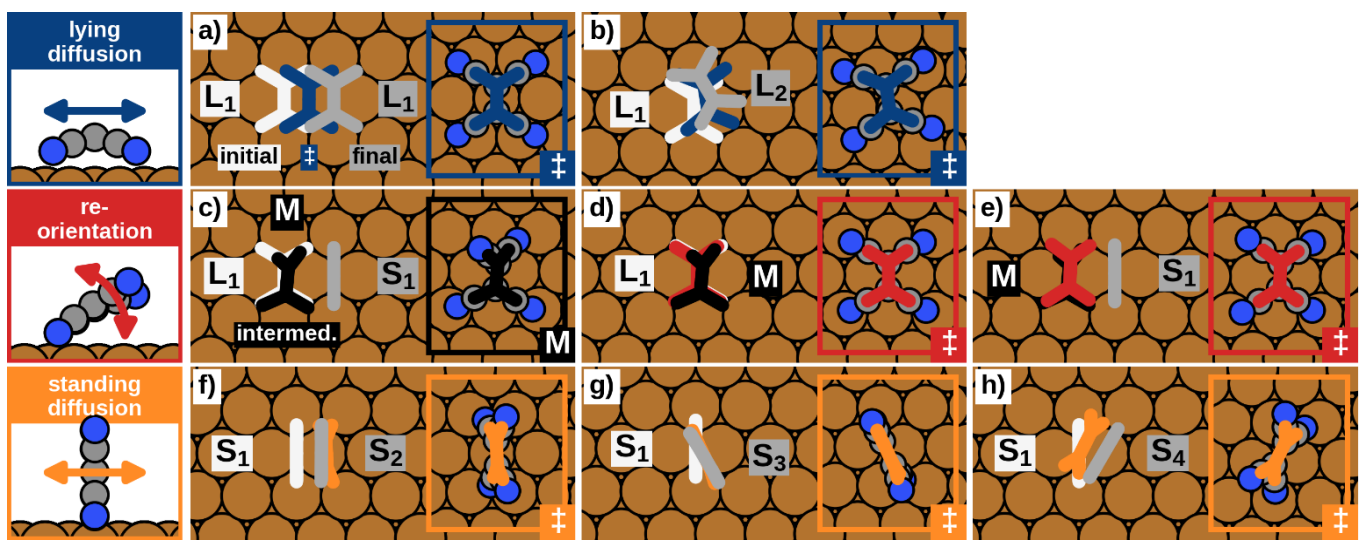

**Figure 3.** Overview of the elementary transition processes of the three different motion regimes "lying diffusion" (a,b), "re-orientation" (c-e) and "standing diffusion" (f-h). In addition to the initial (white) and final (grey) adsorption geometries the positions of the transition states including their specific geometries are provided as well, except for **L₁→S₁** (c), where in-stead the obtained intermediate minimum **M** (black) is shown. The relative positions during the transitions itself are presented in a reduced scheme by omitting the nitrogen atoms.

**Table 1.** Energetics of the elementary transitions: Adsorption energies of the initial ($E^{ini}$), transition ($E^{‡}$) and final state ($E^{fin}$) and the corresponding barriers of the forward ($\Delta E_1^{‡}$) and the reverse ($\Delta E_{-1}^{‡}$) transition.

| Transition | $E^{ini}$ / eV | $E^{fin}$ / eV | $E^{‡}$ / eV | $\Delta E_1^{‡}$ / eV | $\Delta E_{-1}^{‡}$ / eV |
|---|---|---|---|---|---|
| L1→L1 | -2.40 | -2.40 | -1.85 | 0.55 | 0.55 |
| L1→L2 | -2.40 | -2.34 | -1.99 | 0.41 | 0.35 |
| L1→M | -2.40 | -2.20 | -2.10 | 0.30 | 0.10 |
| M→S1 | -2.20 | -1.86 | -1.81 | 0.38 | 0.05 |
| S1→S2 | -2.40 | -1.86 | -1.81 | 0.58 | 0.05 |
| S1→S3 | -1.86 | -1.87 | -1.81 | 0.05 | 0.06 |
| S1→S4 | -1.86 | -1.83 | -1.79 | 0.07 | 0.04 |

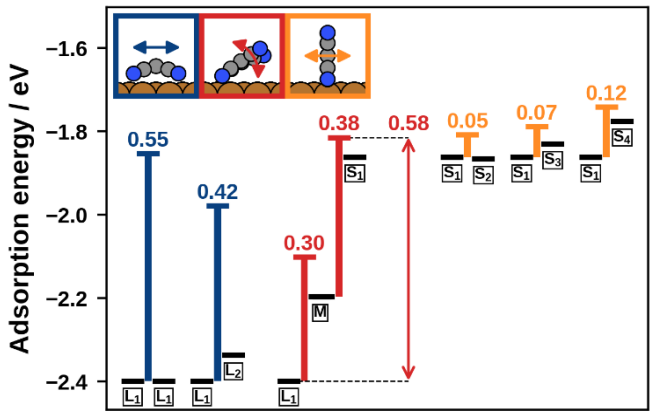

**Figure 4**. Adsorption energies of initial geometries, transition states, and final geometries of the various transitions. The black bars represent the energies of the local minima whereas the colored bars represent the energy barriers of the forward transitions. The arrow denotes the effective barrier for re-orienting from lying to standing TCNE.

Since the **L$_1$→S$_1$** re-orientation process will strongly determine the phase-transition versus growth behavior, we discuss this process in more detail. In Figure 5, the energetic and geometric course of the re-orientation is visualized. In addition to the path (black) where substrate atoms were included in optimizations, also the course (grey) obtained by constraining the substrate during optimizations is shown. Therein, the different shifts of adsorption energies and the change of the energy barrier by 0.2 eV underline that the influence of the substrate is not negligible. In general, the re-orientation occurs in a two-step process: The intermediate minimum **M** is 0.20 eV energetically less beneficial than **L$_1$**. The barrier of **L$_1$→M** is 0.30 eV, and thus smaller than the barrier of **M→S$_1$**, which amounts to 0.38 eV. The effective total barrier for standing up, i.e. the difference between the lowest (**L$_1$**) and highest point (transition state ‡$_B$ in Figure 5) of the pathway, is 0.58 eV. For the reverse transition the molecule needs to overcome only a minute barrier of 0.05 eV between **S₁** and **M**. The backward reaction is complected by overcoming the barrier between **M** and **L₁** (0.10 eV).

In geometric terms, the re-orientation proceeds as follows: The lying TCNE detaches one CN-group, hereafter referred to as "arm", from the surface before reaching the first transition state (‡$_A$). The molecule gains stability again at the intermediate minimum (**M**) by repositioning its opposite arm from the top to the hollow site. After the arm next to the already detached one breaks the second CN-Cu bond, both detached arms come closer to each other, until arriving at the second transition state (‡$_B$). Here the now nearly flat molecule encloses an angle of approximately 30° with the substrate surface. By rotating further into an upright position, the adsorption geometry **S₁** is reached.

It is likely that the re-orientation process of **L₂**→**S₄** follows a similar pathway. However, this transition cannot be rate-limiting for the targeted kinetic trapping, because the difference of the adsorption energies of **L₂** and **S₄** (0.56 eV) is already as large as the barrier of **L₁**→**S₁** (0.58 eV).

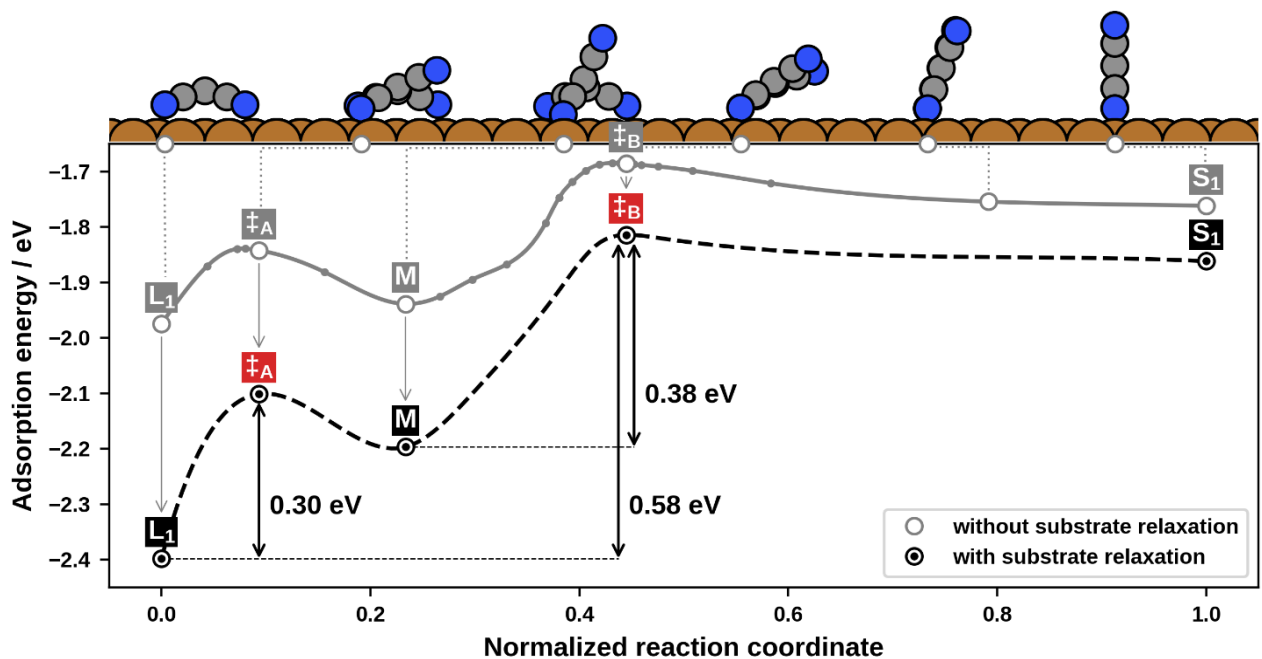

**Figure 5**. Energy evolution while re-orientating from flat-lying to upright-standing position with (grey line) and without (black line) constraining the substrate atoms throughout optimizations. The course with the fixed substrate was sampled by 25 images, whereas for the more accurate description, that includes the influence of the substrate, only the minima and the transition states were re-optimized. The interconnecting, dashed line is only an interpolation between these re-optimized points. In addition, geometries of characteristic positions are provided in the side view for the initial (**L₁**), intermediate (**M**) and final (**S₁**) minima, as well as the two transition states ‡$_A$ and ‡$_B$.

## 2.2 Transition rates

To determine under which conditions the re-orientation of TCNE molecules can be prevented, while still allowing for growth of the flat-lying structures, we need to obtain temperature-dependent transition rates by utilizing the energy barriers. We assume that in a hypothetical physical vapor deposition (PVD) experiment an ordered flat-lying structure can form as long as the temperature is sufficiently high for the molecules to readily diffuse. The speed at which this structure grows is then limited by the available material. In a PVD experiment, this is given by the rate at which TCNE molecules are deposited onto the substrate. Furthermore, it is plausible to assume that the structure becomes (kinetically) stabilized once it reaches mesoscopic dimensions or becomes buried under a significant amount of material, i.e. once the deposited TCNE is several layers thick. In other words, if the growth to multilayers occurs faster than the time required for even a single TCNE to re-orient into an upright position, we assume to have kinetically trapped the flat-lying structure. In short, we need to find a temperature range where a) diffusion of molecules is much faster and b) the re-orientation is considerably slower than a given deposition rate.

We can calculate temperature-dependent transition rates $k(T)$ utilizing the harmonic transition state theory (see Methods for details) with energy barriers $\Delta E^{\ddagger}$ from Table 1 and attempt frequencies $A$, as provided in Table 2:

$$k(T) = A\, e^{-\frac{\Delta E^{\ddagger}}{k_B T}} \qquad \text{(Equation 1)}$$

Since our goal is to prevent the re-orientation of individual molecules to the upright-standing position, we want to discuss a joint process of the re-orientation ($L_1 \rightarrow S_1$) rather than the separate elementary transitions $L_1 \rightarrow M$ and $M \rightarrow S_1$. Thus, we assign an effective barrier of 0.58 eV (see Figure 5) to the joint process of standing up. For the lying down $S_1 \rightarrow M$ is the decisive step (barrier of 0.05 eV).

Apart from energy barriers, rates are determined by the attempt frequencies of the transitions. In principle, attempt frequencies are the vibration frequencies in direction of the reaction coordinate. Within harmonic transition state theory, they are explicitly obtained from the stable vibration frequencies of the initial and the transition states (for details see Methods). As Table 2 shows, the attempt frequencies are very different for different processes, covering five orders of magnitude. The joint process of lying down ($S_1 \rightarrow L_1$) has with 8.8 x $10^{11}$ Hz the lowest attempt frequency, while the largest is obtained for the $L_1 \rightarrow L_1$ diffusion with 2.0 x $10^{15}$ Hz. In total, attempt frequencies for the lying diffusion are by up to three orders of magnitude larger than the ones of the standing diffusion. For the re-orientation process, attempt frequencies of the standing up are at least ten times larger than the attempt frequencies of falling-over.

**Table 2.** Attempt frequencies obtained by means of harmonic transition state theory. The subscripts 1 and -1 denote forward and reverse transitions, respectively.

|  | L$_1$→L$_1$ | L$_1$→L$_2$ | L$_1$→S$_1$ | S$_1$→S$_2$ | S$_1$→S$_3$ | S$_1$→S$_4$ |
|---|---|---|---|---|---|---|
| *A$_1$* / Hz | 2.0 x 10$^{15}$ | 2.9 x 10$^{14}$ | 5.0 x 10$^{13}$ | 1.7 x 10$^{12}$ | 1.0 x 10$^{12}$ | 1.7 x 10$^{13}$ |
| *A$_{-1}$* / Hz | 2.0 x 10$^{15}$ | 1.3 x 10$^{14}$ | 8.8 x 10$^{11}$ | 4.6 x 10$^{12}$ | 1.2 x 10$^{12}$ | 1.3 x 10$^{13}$ |

Using these attempt frequencies, the temperature-dependent transition rates are calculated according to Equation 1 and shown in Figure 6. Before explicitly investigating the rates of the single processes, we discuss at which rates we can consider transitions to be suppressed within the growth process. Within PVD experiments, thin films (i.e., multilayers) of organic materials are typically deposited within minutes to (at most) days. This corresponds to deposition rates $k_{dep}$ of about 1 monolayer per minute to 1 monolayer per day. Based on the obtained transition rates (Figure 6), we can identify at which temperatures individual processes occur much more slowly than the deposition process itself. In other words: We can consider single processes as suppressed if transition rates $k < k_{dep}$ are enforced. For the present discussion, we propose a target transition rate $k$ of 1 transition per day (10$^{-5}$ transitions per second), as indicated by the black line. The temperatures required to reach this target transition rate are stated in Table 3. For a concise representation, only the limiting transitions, i.e. the transitions with the highest rates within a class, of the lying diffusion (**L$_1$→L$_2$**), the standing diffusion (**S$_1$→S$_2$**), the standing-up (**L$_1$→S$_1$**) and the lying-down (**S$_1$→L$_1$**) are plotted. In the Supporting Information a visualization of the rates of all transitions is provided, as well as a detailed uncertainty discussion including root mean square uncertainty estimates of the obtained suppression temperatures. Summarizing the outcome, we estimate the uncertainty of the suppression temperatures of the lying diffusion to ≈ 30 K, whereas the one for standing up is with ≈ 40 K the largest error estimate. For all other transitions the uncertainty is ≈ 10 K.

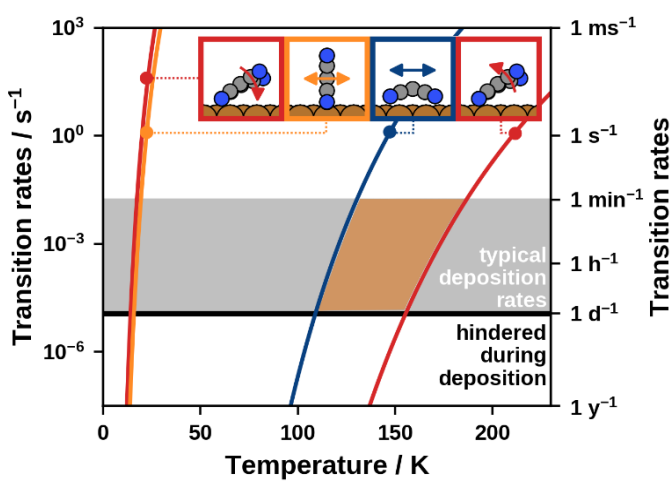

**Figure 6.** Limiting rates of lying diffusion **L$_1$→L$_2$** (blue), re-orientation **L$_1$↔S$_1$** (red) and standing diffusion **S$_1$→S$_2$** (orange) as function of the process temperature. The range of typical deposition rates is marked by the grey area and the limiting tran-

sition rate for hindering distinct processes during deposition is indicated by the black line at 1 transition per day. Within the former, the area where the monolayer of lying molecules is kinetically trapped against re-orientation is highlighted.

**Table 3.** Estimated temperatures for suppression $T$ computed for transition rates of 1 transition per day. The subscripts 1 and -1 refer to forward and reverse transitions, respectively.

|             | $L_1 \rightarrow L_1$ | $L_1 \rightarrow L_2$ | $L_1 \rightarrow S_1$ | $S_1 \rightarrow S_2$ | $S_1 \rightarrow S_3$ | $S_1 \rightarrow S_4$ |
|-------------|-----|-----|-----|-----|-----|-----|
| $T_1$ / K   | 140 | 110 | 160 | 20  | 20  | 30  |
| $T_{-1}$ / K | 140 | 90  | 10  | 20  | 10  | 10  |

As shown in Figure 6, the diffusion of standing TCNE along single symmetry axes ($S_1 \rightarrow S_2$) and the lying-down ($S_1 \rightarrow L_1$) are the fastest processes and obtain similar rates. Therefore, standing molecules might not diffuse over long distances before falling over again. For temperatures above 100 K all processes of standing diffusion and the lying-down occur on sub-ns timescales. The diffusion of the flat-lying molecules freezes out at temperatures below 110 K. This temperature is relatively high due to concurrent relaxations of the substrate, specifically a "pulling out" of Cu atoms bonded to the nitrogen atoms by 0.2 Å. This relaxation lowers the barrier by 0.42 eV. Without it, the diffusion would freeze out at temperatures below 60 K. For temperatures above 140 K, the lying diffusion proceeds on the order of seconds, which increases to the order of µs at room temperature. Finally, the process of standing up ($L_1 \rightarrow S_1$) is the slowest. For temperatures below 160 K we estimate a rate of less than one transition per day. At room temperature it still occurs very efficiently (millisecond timescale). Here, we remind the reader that this joint process is, indeed, a two-step process. For the sake of completeness, the rates for these two elementary processes are provided in the Supporting Information (Figure S7).

Based on these results, we predict that, in the temperature range of 110 to 160 K, the standing up of individual molecules can be suppressed while molecules can still diffuse.

For a process temperature of 140 K, lying molecules diffuse by a rate of ≈ 0.2 transitions per second. This should be sufficiently high to ensure a diffusion mobility that allows building at least a full lying monolayer within the deposition time. For lower temperatures, growth of ordered structured is likely to be inhibited by random aggregation of impinging molecules. For 140 K, the rate of standing up is 0.3 transitions per year. Once they are standing, the molecules fall down again within nanoseconds. Therefore, it is unlikely that standing seeds are created during the growth process, provided that deposition rates are low enough to avoid aggregation. For this regime, detailed knowledge of the influence of standing seeds on the stability of standing molecules becomes dispensable.

At the first sight, this prediction is qualitatively not consistent with the experiments of *Erley and Ibach*, who observed both, standing and lying molecules for deposition at 100 K.[20,37] While the formation of lying seeds is covered within the estimated uncertainty, standing seeds should not form regarding to our predictions. This contradiction might either result from a) a temperature increase in the experiment during the highly exothermal deposition process or b) from the reduction of the energy barrier caused by collaborative effects.

For coverages that exceed the one of the favored flat-lying structure, i.e. when a second layer is created, we assume that the re-orientation of the whole monolayer can be suppressed as well for temperatures below 160 K. At higher dosages, the re-orientation rate will increase, but by depositing further layers fast enough until the layer thickness reaches a mesoscopic scale, we expect to prevent the re-orientation of the whole first layer. To check, whether this assumption holds true, detailed investigations about the intermolecular and interlayer processes will need to be performed in the future.

## 3. CONCLUSION

To propose experimental conditions that prevent the re-orientation of flat-lying molecules in the first adsorbate layer to the thermodynamically favored upright-standing positions, we studied kinetic processes of tetracyanoethylene (TCNE) molecules on a Cu(111) surface. Utilizing the nudged elastic band method and the harmonic transition state theory, energy barriers and transition rates were obtained for the diffusion of lying and standing TCNE molecules, as well as for the re-orientation between these two positions. The most dominant and thus limiting re-orientation process turned out to advance in two steps, exhibiting an effective energy barrier of 0.58 eV for standing up and 0.05 eV for lying down. Based on the obtained rates, we estimate that for temperatures above 110 K a sufficiently high diffusion mobility is ensured, which further allows the formation of an ordered monolayer of flat-lying TCNE. While our investigation reveals that individual molecules can be prevented from standing up for temperatures below 160 K, this finding offers an initial indication for the behavior of the whole monolayer. Determining this temperature more precisely and assessing how long the first layer remains kinetically trapped upon deposition of further layers will require further studies on the intermolecular and interlayer processes. Nevertheless, this work constitutes a first step towards fully understanding transition processes of organic thin films.

## 4. METHODS

In this work, the sampling of the potential energy surface is conducted within the framework of Kohn-Sham density functional theory as implemented in the software package FHI-aims.[38] We use the PBE[39] functional and the TS$^{surf}$ dispersion correction.[40] We apply the repeated slab approach with periodic boundary conditions in all three dimensions. Unit cell heights of 68 Å ensure vacuum heights of at least 50 Å between two consecutive slabs. Hereby a dipole correction[41] is used to electrostatically decouple the replicas in z-direction. The TCNE molecules are placed on a substrate consisting of 7 copper layers, whose lattice constant (of 2.55 Å) was obtained by a Birch-Murnaghan fit. The band structure is sampled using a generalized Monkhorst-Pack grid[42–44] with a spacing of $\Delta k = \frac{2\pi}{8}$ nm$^{-1}$. A Gaussian broadening of 0.1 eV is applied to all states.

FHI-aims employs numeric atom-centered basis functions. In this work, we use the "tight" default settings for C and N. The three uppermost Cu layers are also treated with the "tight" species defaults, whereas the residual four layers are treated with "light" settings to save computational time. This is described in detail in the supplementary information of a previous publication[20], where identical DFT settings are used.

The convergence criteria of the SCF-procedure were set to 1 x $10^{-2}$ eÅ$^{-3}$ for the charge density, 1 x $10^{-5}$ eV for the energy and 1 x $10^{-3}$ eV Å$^{-1}$ for the forces.

To achieve converged adsorption energies (within ≈ 20 meV), 6x6 Cu super cells are required.

The resulting energies correspond to electronic energies of the whole system $E_{sys}$ at zero Kelvin. Adsorption energies $E_{ads}$ are determined according to Equation 1, where $E_{mol}$ is the energy of a relaxed molecule in the gas phase and $E_{sub}$ the energy of the pre-relaxed substrate, as used in the slab.

$$E_{\text{ads}} = E_{\text{sys}} - E_{\text{mol}} - E_{\text{sub}} \qquad \text{(Equation 2)}$$

By this definition, more favored adsorption geometries are connected to more negative energies.

While molecular dynamics simulations are the standard method for kinetic studies, its application isn't affordable for the investigated system and purpose. As timesteps for similar systems are typically in the order of femtoseconds or less, the available simulation time is not sufficient for reliably escaping basins of a wide area of attraction to measure barriers and rates of e.g. processes like re-orientations. The computational cost is further increased by sampling the potential energy surface on the level of density functional theory, which is necessary to capture the underlying chemistry. Even though advances in accelerating sampling of rare events[45–47] have been made, we decided on using transition path sampling methods instead. Transition rates between single adsorption minima are provided *via* harmonic transition state theory, whereas energy barriers themself are determined beforehand with a transition state search method.

For the transition state search, the climbing image nudged elastic band (CI-NEB) method[21,22] augmented with the Fast Inertial Relaxation Engine (FIRE) optimizer[48] is applied. The workflow can be summed up as the following: The transition path is initialized in a 4x4 super cell between the selected pair of minima with up to five images using the image dependent pair potential (IDPP) method[49] as shipped by the software package ASE.[50] After several iterations, only the images with the highest energies and/or forces are updated for efficiency reasons. Once the NEB force of the image with the highest energy drops below 0.01 eV/Å, further images are inserted and converged to verify that the highest barrier along the path is found. All transition paths are sampled by 7 to 25 images with a maximum residual NEB force of 0.05 eV/Å, whereas the NEB forces of all transition states are below ≈ 0.01 eV/Å. In addition, all transition states are re-optimized in a 6x6 super cell where the atoms of the two uppermost copper layers are set unconstrained: At first only the substrate is allowed to relax, while afterwards the transition state is re-optimized by unconstraining both, the adsorbate and the substrate to NEB forces < 0.01 eV/Å. The transition state is, per definition, a first order saddle point, where the Hessian exhibits exactly one instable eigenmode that corresponds to a negative curvature or eigenfrequency.

Numerical vibrational analyses were performed (Γ-point, 4x4 super cell, fixed substrate) for all minima and transition states, in order to a) assure that the transition states have only one negative frequency, and b) to obtain the attempt frequencies required for the harmonic transition state theory. Displacements of 0.01 Å are applied for computing Hessians. In the Supporting Information, we explain why Hessians are symmetrized and how additional instable frequencies at transition states and minima are treated.

The harmonic transition state theory[25,26] enables determining the transition rates *k* as stated in Equation 4.

$$k = A\, e^{-\frac{\Delta G^{\ddagger}}{k_B T}} \quad \text{with} \quad A = \frac{\prod_{i=1}^{3N} \nu_i^{\text{ini}}}{\prod_{i=1}^{3N-1} \nu_i^{\ddagger}} \quad \text{and} \quad \Delta G^{\ddagger} = G^{\ddagger} - G^{\text{ini}} \qquad \text{(Equation 4)}$$

Therein, *k* is the product of the harmonic attempt frequency *A* and the Boltzmann-factor containing the Gibbs free energy barrier *ΔG*‡ and the temperature *T*. *ΔG*‡ is the difference of the Gibbs free energy of the transition state (superscript ‡) and the initial state (superscript ini), and in general also depends on the electronic energy, the temperature, the pressure and the unit cell size. In this study, however, the unit cell size and the number of adsorbate molecules per unit cell stays unchanged for all transitions which reduces the dependency of *ΔG*‡ to the pure electronic energy barrier *ΔE*‡ (see Supporting Information for details). *A* is the ratio of the products of the stable vibration frequencies $\nu_i$ at the initial and the transition state.

**ASSOCIATED CONTENT**

Supporting Information

The Supporting Information contains numerical settings of the density functional calculations, energy courses of all elementary transitions, additional information regarding the computed rates and a detailed uncertainty discussion of the temperatures required to suppress distinct processes.

Data Availability

The computed transition paths are available in the NOMAD database (https://nomad-lab.eu/). For the review process, transition paths can be found when searching "authors=Anna Werkovits" in the NOMAD search bar (https://nomad-lab.eu/prod/rae/gui/search). DOIs of the data sets will be provided prior to publication.

**AUTHOR INFORMATION**

Corresponding Author

Oliver Hofmann - Institute of Solid State Physics, NAWI Graz, Graz University of Technology, Petersgasse 16, 8010 Graz, Austria; orcid.org/0000-0002-2120-3259; Phone: +43 316873 8964; Email: o.hofmann@tugraz.at

**ACKNOWLEDGMENTS**

We acknowledge fruitful discussions with F. Calcinelli, B. Ramsauer, R. Berger and R. Steentjes. Funding through the START project of the Austrian Science Fund (FWF): Y1175-N36 is gratefully acknowledged. Computational results have been achieved using the Vienna Scientific Cluster (VSC).

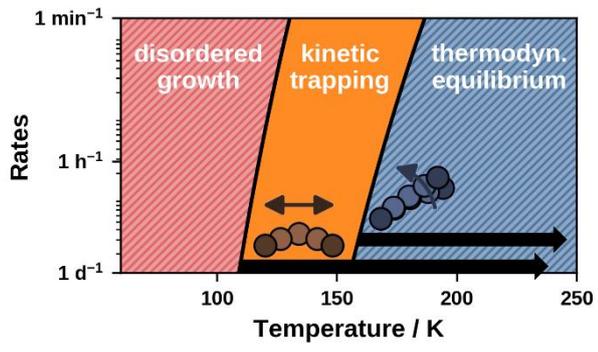

# Supporting Information

for

"Towards targeted kinetic trapping of organic-inorganic interfaces: A computational case study"

Anna Werkovits, Andreas Jeindl, Lukas Hörmann, Johannes J. Cartus, Oliver T. Hofmann*
Institute of Solid State Physics, TU Graz, NAWI Graz, Petersgasse 16/II, 8010 Graz, Austria

# Table of Content

# 1 Thermodynamically stable monolayers of TCNE/Cu(111)

As predicted by Egger[1] et al., there are two favorable structures for different coverage ranges: For low coverages, the flat-lying structure from Figure S1 is thermodynamically the most favorable, whereas for increasing coverages the herringbone structure with upright-standing molecules (Figure S2) becomes superior.

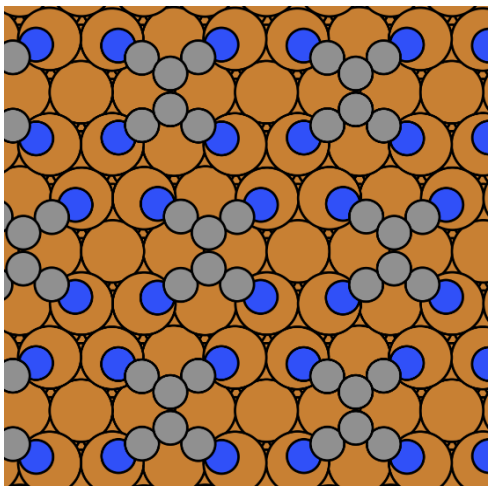
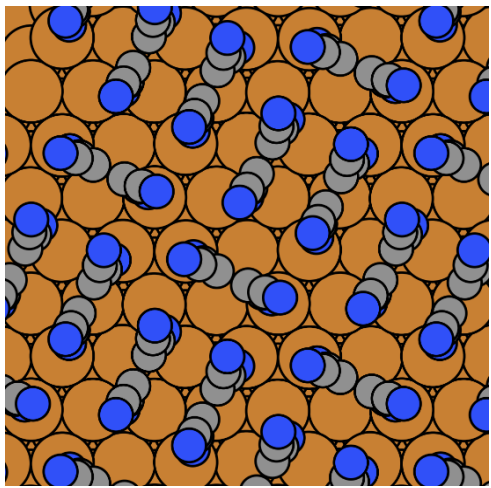

***Figure S1:*** *Flat-lying monolayer*      ***Figure S2:*** *Upright-standing monolayer*

# 2 Transitions
## 2.1 Example for identification of non-elementary transition: $S_1 \rightarrow S_1$ transition

In Figure S3 the transition $S_1 \rightarrow S_1$ is initialized in linear fashion. After some iterations of the nudged elastic band method, we see in Figures S4 and S5 that the central image converged towards S2. Therefore, splits in the equivalent transitions $S_1 \rightarrow S_2$ and $S_2 \rightarrow S_1$.

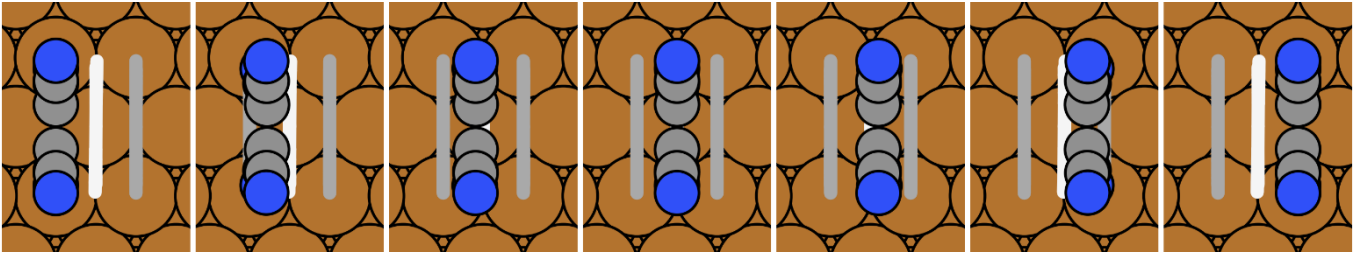

***Figure S3:*** *Initialized $S_1 \rightarrow S_1$ transition*

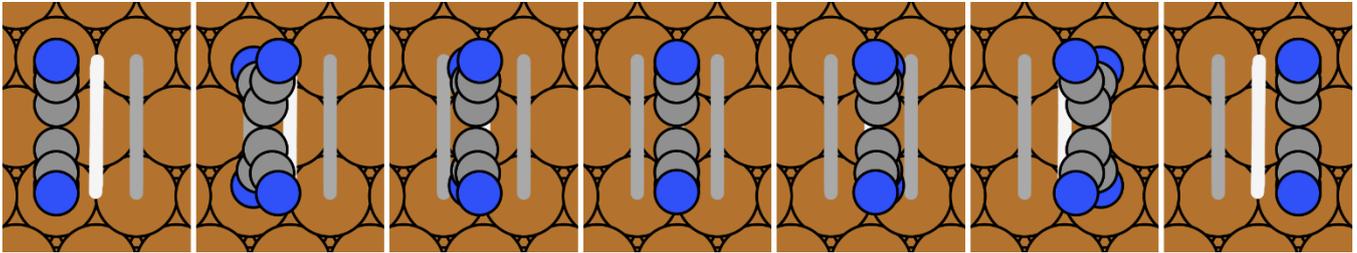

***Figure S4:*** *$S_1 \rightarrow S_1$ transition after 74 iterations of the nudged elastic band method*

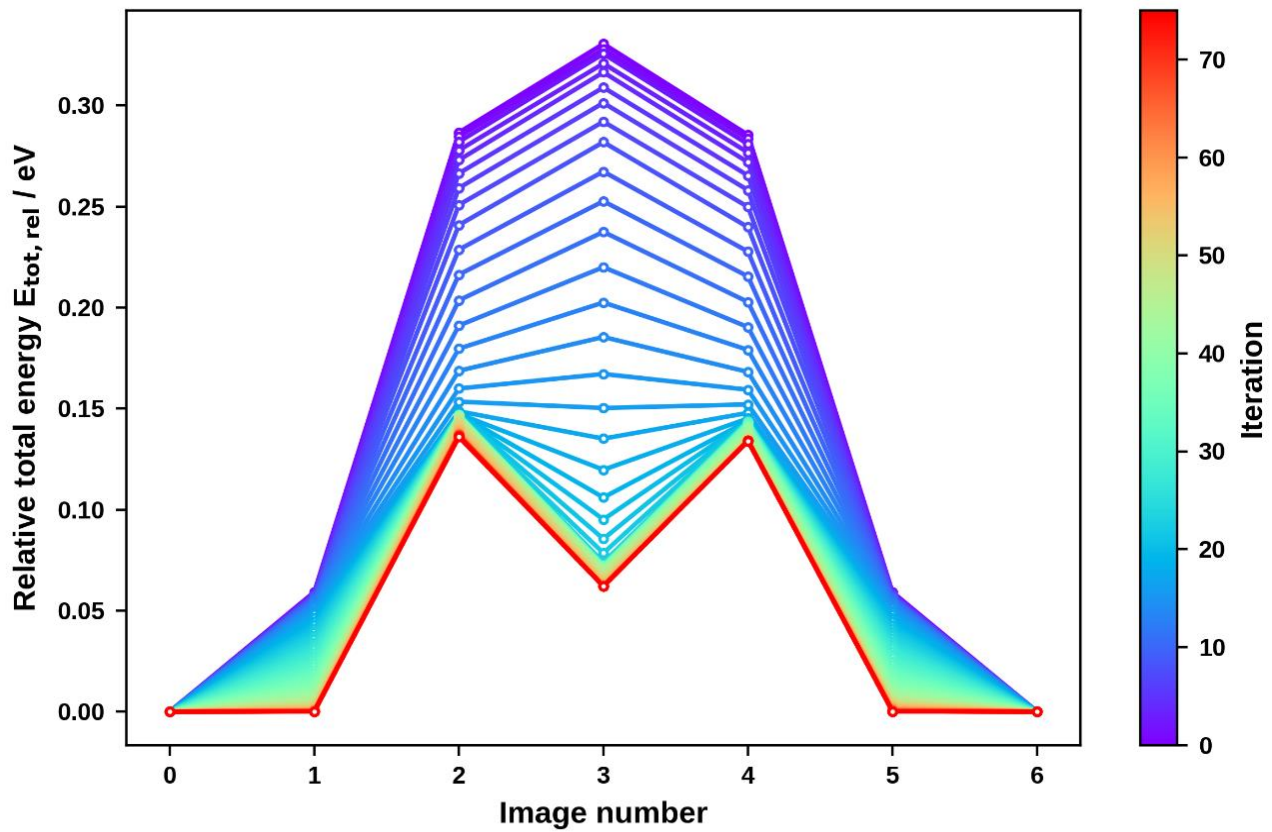

***Figure S5:*** *Energy evolution of the $S_1 \rightarrow S_1$ transition after applying the nudged elastic band method*

## 2.2 Elementary transitions

Figure S6 depicts changes in adsorption energy in dependence of the normalized reaction coordinate for all transitions. In addition, the influence of including substrate relaxations to the optimizations is visualized.

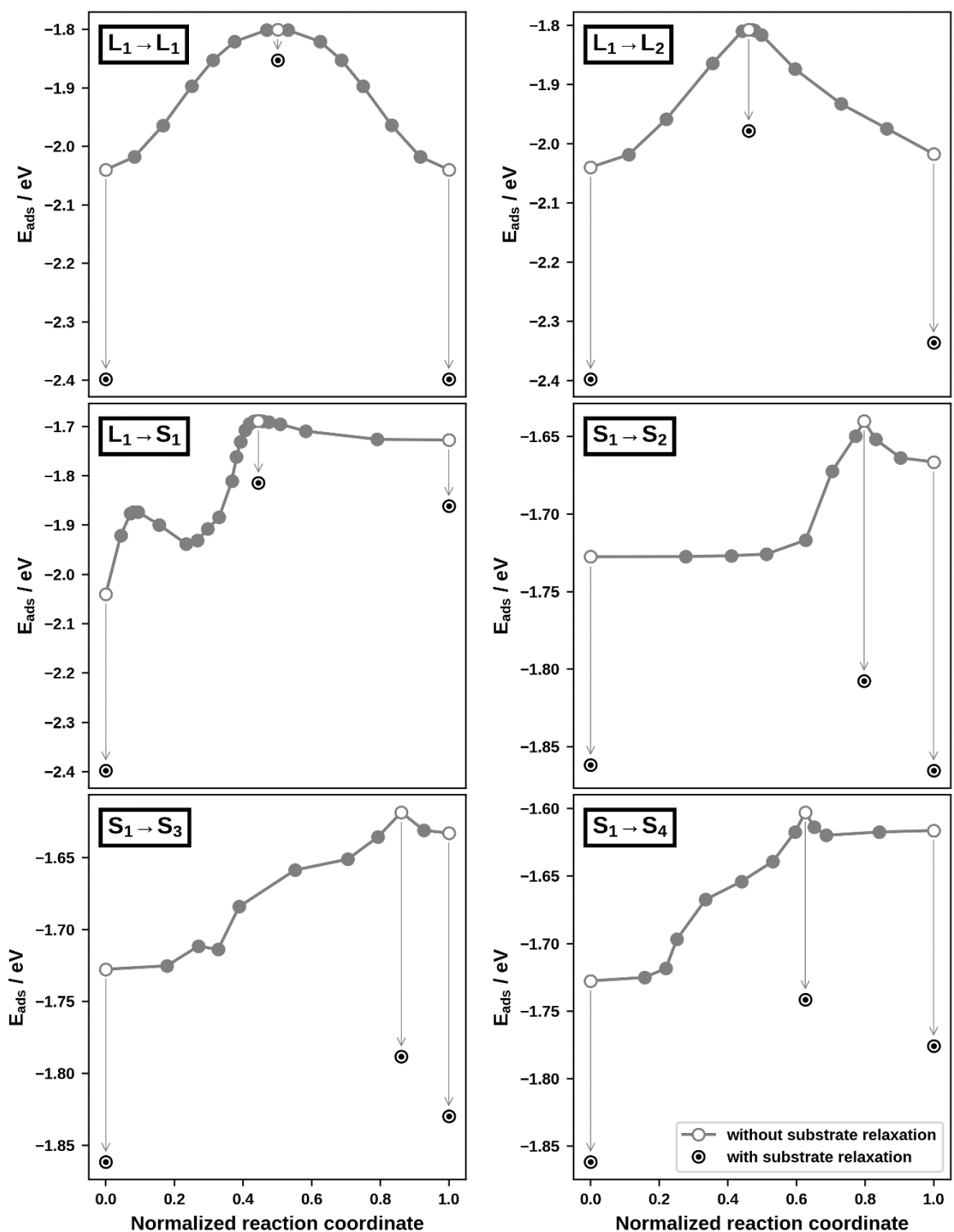

*Figure S6:* Sampled minimum energy paths described by the change of adsorption energy along the normalized reaction coordinate. The paths in grey are simulated in a 4x4 super cell and a fixed substrate, whereas the black dots show the change after re-optimization in a 6x6 super cell by including substrate relaxation.

# 3 Transition Rates

## 3.1 Treatment of the Gibbs free energy

We apply *ab initio* thermodynamics[2] to include finite temperature effects to the energies as obtained via density functional theory. For modelling growth experiments as closed systems at constant temperature and pressure, the Gibbs free energy is the relevant thermodynamic potential. In Equation 1, the Gibbs free energy of adsorption is constructed from the adsorption energy $E_{ads}$ and the contribution of the chemical potential $\mu_{ads}$, while mechanical work, configuration entropy and vibration enthalpy are neglected, as it is commonly done in literature.[2]

$$G_{ads} = E_{ads} - \mu_{ads} N_{ads} \qquad \text{(Equation 1)}$$

$N_{ads}$ is the number of molecules adsorbed per unit cell, whereas $\mu_{ads}$ is the chemical potential of the molecule in the gas phase, that depends on temperature and pressure. $\mu_{ads}$ is obtained in the ideal gas approximation by simply using translational and rotational contributions as provided in the thermochemistry package distributed within ASE[3].

For our case, all unit cells are identical and include only one adsorbate, i.e. $N_{ads}$ = 1. Therefore, $G_{ads}$ directly reduces to $E_{ads}$.

## 3.2 All transition rates

As only the joint processes and not the elementary transitions of the re-orientation are relevant for predicting process conditions for kinetically trapping the first layer of flat-lying molecules, the attempt frequencies of the two elementary steps of the re-orientation process are not stated in the main manuscript. Thus, these are provided in Table S1.

**Table S1:** *Attempt frequencies obtained by means of harmonic transition state theory. The subscripts 1 and -1 denote forward and reverse transitions, respectively.*

|  | $L_1 \rightarrow M$ | $M \rightarrow S_1$ |
|---|---|---|
| $A_1$ / Hz | 1.7 x $10^{13}$ | 5.0 x $10^{13}$ |
| $A_{-1}$ / Hz | 9.0 x $10^{12}$ | 8.8 x $10^{11}$ |

In Figure S7, also the transition rates of all elementary transitions and joint processes are visualized. Figure S8 shows the same content, but in a wider range.

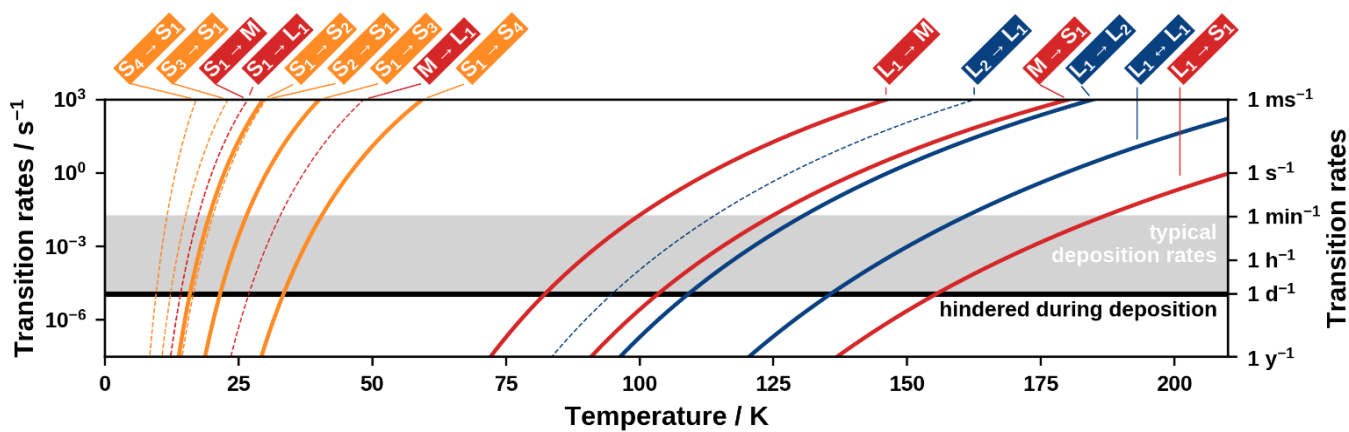

*Figure S7: Transition rates in dependence of temperature including the two elementary transitions of the re-orientation*

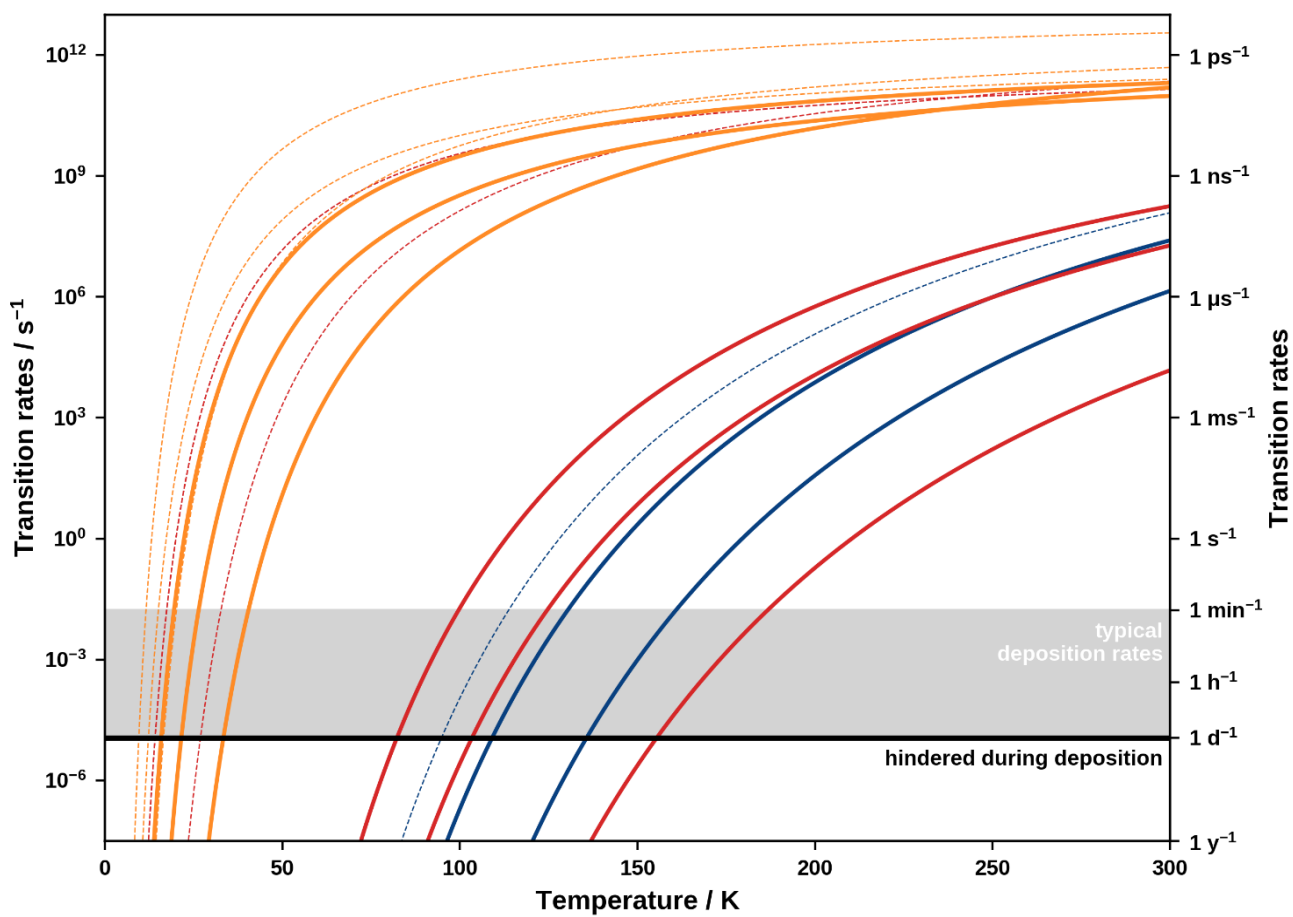

*Figure S8: Transition rates in dependence of temperatures in an extended range (labeling according to Figure S7)*

## 3.3 Joint process of re-orientation

Since our goal is to prevent the re-orientation of individual molecules to the upright-standing position, we want to discuss a joint process of the re-orientation rather than the separate elementary processes.

For the joint process of standing-up ("$L_1 \rightarrow S_1$") the rate determining transition state is the one of step $M \rightarrow S_1$. This is evident in Figure S9. Thus, we assign an effective barrier of 0.58 eV to the joint process of standing-up, as indicated in Figure 5 of the main manuscript. This is constituted of the sum of the pure difference in adsorption energies of $L_1 \rightarrow M$ (0.20 eV) and the energy barrier of $M \rightarrow S_1$ (0.38 eV). This description is formally valid under the assumption that $L_1$ and M are in a pre-equilibrium. In addition, the attempt frequency of $M \rightarrow S_1$ is applied ($5.0 \times 10^{13}$ s$^{-1}$).

For the joint process of lying down ("$S_1 \rightarrow L_1$"), we are especially interested in the step $S_1 \rightarrow M$. Therefore, we employ the $M \rightarrow L_1$ barrier of 0.05 eV and its attempt frequency ($8.8 \times 10^{11}$ s$^{-1}$).

## 4 Uncertainty discussion

In general, there are several sources of uncertainty or errors arising from the applied DFT functional, geometric constraints, the transition state search, the transition state theory and the thermodynamic description.

Here, we discuss how these uncertainties influence the uncertainty of the temperatures that are required for suppression of single elementary or joint processes. In detail, we include the uncertainty of the transitions state and especially the influence from the vibrational frequencies.

### 4.1 Transition states and energy barriers

The quality of the obtained transition states is mainly influenced by the accuracy of the DFT calculation ($\approx$ 20 meV), as well as by the performance of the nudged elastic band method (NEB) method. In general, the geometric path used for the initialization of the NEB method can influence the resulting minimum energy path. Therefore, the path converges either to the global or only to a local solution. Checking the stability of the resulting transition paths for different initial paths is not affordable. Besides this influencing factor, the sampling resolution along the obtained path is important as well: To ensure that the highest barrier of the resulting path is captured, a plausible resolution hast to be achieved. In addition, the transition state should be sampled as accurately as possible for a trustworthy measure of the energy barrier and the vibrational frequencies required for computing transition rates. Trading off the accuracy with the involved cost, we tried to push the residual NEB forces $F_{NEB}$ on the transition state below 0.01 eV Å$^{-1}$. The resolution of the chain of images is expressed via the maximal distance $\Delta x$ atoms of the adsorbate exhibit between two neighbored images. For a resolution of $\Delta x = 1$ Å, this would refer to a maximal uncertainty of the transition state energy of $\Delta E = F_{NEB} \cdot \Delta x = 10$ meV.

In Table S2, the number of images per run, the maximal residual NEB forces of the whole chain and especially for the obtained transition state are provided, as well as the resolution at the transition state. Residual forces at the transition states of < 0.01 eV Å$^{-1}$ could be obtained. The resolution at the transition state is around 0.05 to 1.00 Å. When accounting for the whole chain of images for all calculated transitions, the resolution ranges from 0.04 to 1.92 Å. Here, we remind the reader that these calculations have been conducted in a 4x4 super cell.

*Table S2:* Force accuracies and information about resolution of the conducted NEB runs in a 4x4 super cell. The number of sampled images is $n_{images}$. The absolute value of the maximal NEB force acting on a single atom of TCNE along the sampled transition path is denoted as $|F|_{NEB,max}$, whereas $|F|_{NEB,max,TS}$ states the maximum force acting on the transition state. The resolution of the sampled images is expressed via the maximum distance atoms of the adsorbate exhibit between two neighbored images. Here, $\Delta x_{max,pre\ TS}$ and $\Delta x_{max,post\ TS}$ provide the resolution at the transition state.

|  | $n_{images}$ | $|F|_{NEB,max}$ / eV Å$^{-1}$ | $|F|_{NEB,max,TS}$ / eV Å$^{-1}$ | $\Delta x_{max,pre\ TS}$ / Å | $\Delta x_{max,post\ TS}$ / Å |
|---|---|---|---|---|---|
| $L_1 \rightarrow L_1$ | 15 | 1.8E-02 | 3.5E-03 | 0.10 | 0.09 |
| $L_1 \rightarrow L_2$ | 13 | 3.0E-02 | 2.0E-03 | 0.10 | 0.05 |
| $L_1 \rightarrow M$ | 7 | 1.4E-02 | 4.8E-03 | 0.25 | 1.00 |
| $M \rightarrow S_1$ | 19 | 4.6E-02 | 4.3E-03 | 0.09 | 0.10 |
| $S_1 \rightarrow S_2$ | 11 | 3.9E-02 | 1.2E-02 | 0.23 | 0.22 |
| $S_1 \rightarrow S_3$ | 11 | 3.0E-02 | 2.7E-03 | 0.36 | 0.30 |
| $S_1 \rightarrow S_4$ | 13 | 3.9E-02 | 6.9E-03 | 0.27 | 0.29 |

In Section 4.2.2 Figure S10 we will show that the uncertainty of the transition state energy is rather < 1 meV than < 10 meV for the 4x4 super cell. Nevertheless, the transition states have been re-optimized in a 6x6 super cell in two steps. Firstly, the atoms of the two uppermost copper layers of the transition state were relaxed. Secondly, a single-image NEB run was conducted for an unconstrained motion of both, the substrate and the adsorbate atoms. Except for the $L_1 \rightarrow L_1$ transition, the convergence threshold of 0.01 eV Å$^{-1}$ had already been reached with the first step. Under the assumption that uncertainty of the transition state has not changed after the re-optimizations, we assign to the energy barriers an uncertainty of 40 meV, i.e. the maximum error one obtains by calculating differences of adsorption energies with an uncertainty of 20 meV.

## 4.2 Vibration frequencies

For the computation of vibrational analyses at the obtained minima and transition states displacements of 0.01 Å are utilized. In addition, two corrections regarding the symmetry of the hessian and the quality of low and instable frequencies are applied as discussed in Section 4.2.1 and 4.2.2, respectively. Furthermore, relative uncertainties of vibration frequencies are estimated as well in Section 4.2.3. We remind the reader that these calculations have been conducted in a 4x4 super cell.

### 4.2.1 Asymmetricity

As Hessians are obtained in numerically, they are not perfectly symmetric. Therefore, they are symmetrized. These asymmetries can be measured by calculating the root mean square value (rms) of the difference of the original and the symmetrized Hessian, as well as by the maximal absolute value of this difference (max). Accordingly, values of 0 eV Å$^{-2}$ refer to ideal symmetric Hessians. Tables S3 and S4 show that rms of the asymmetricity ranges from 0.7 to 1.4 eV Å$^{-2}$, whereas the maximal values lie between 0.1 and 0.3 eV Å$^{-2}$ for minima and transition states.

*Table S3: Asymmetry of Hessian matrices for the minima.*

| Minimum | $L_1$ | $L_2$ | M | $S_1$ | $S_2$ | $S_3$ | $S_4$ |
|---|---|---|---|---|---|---|---|
| Rms / eV Å$^{-2}$ | 0.87 | 0.78 | 1.40 | 0.83 | 0.87 | 0.99 | 1.44 |
| Max / eV Å$^{-2}$ | 0.10 | 0.09 | 0.27 | 0.09 | 0.15 | 0.14 | 0.34 |

*Table S4: Asymmetry of Hessian matrices of the transition states.*

| Transition state | $L_1 \rightarrow L_1$ | $L_1 \rightarrow L_2$ | $L_1 \rightarrow M$ | $M \rightarrow S_1$ | $L_1 \rightarrow S_1$ | $S_1 \rightarrow S_2$ | $S_1 \rightarrow S_3$ | $S_1 \rightarrow S_4$ |
|---|---|---|---|---|---|---|---|---|
| Rms / eV Å$^{-2}$ | 0.71 | 1.10 | 1.07 | 1.07 | 1.07 | 0.77 | 1.23 | 1.00 |
| Max / eV Å$^{-2}$ | 0.08 | 0.15 | 0.16 | 0.10 | 0.10 | 0.11 | 0.17 | 0.17 |

### 4.2.2 Quality of transition states and minima

After the NEB run converges to the defined convergence criteria, we perform vibrational analyses to ensure that the obtained transition state is indeed a 1$^{st}$ order saddle point. In theory, the vibrational analysis must have exactly one instable (negative) vibration frequency. Similarly, minima must obtain only stable (positive) vibration frequencies. In practice, sometimes small, but positive, frequencies are erroneously identified as additional instable frequencies. This is a common issue inherent in density functional theory originating from numerical integration on a finite k-point grid.[4] In order to check if the unwanted negative frequencies result from this issue or if the sampled transition state (or minimum) is a higher order saddle point, the potential energy surface is distinctly sampled along all instable vibration modes. In detail, displacements (perturbations) along these modes in the range of ± 0.1 Å are applied. The displacements are defined as the maximal displacement of the atoms in this vibration mode.

As schematically visualized in Figure S9, this easily allows then to compare the harmonically approximated vicinity of the potential energy surface along the vibration mode (as probed by the vibration analysis) with the directly calculated energy course.

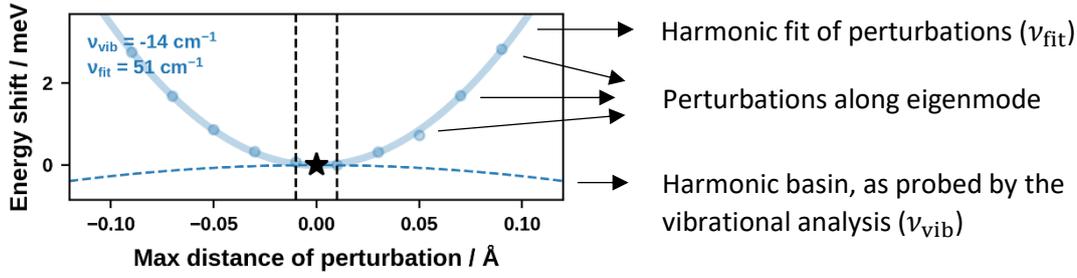

**Figure S9:** *Example for an erroneously obtained instable vibration frequency.*

By checking the consistency of the resulting energy course (minimum, maximum etc.) it becomes clear whether these negative modes occurred from the mentioned numerical inaccuracy or if the instable frequency correctly represents an extremum with a negative curvature. In case of the latter, the transition state search or geometry optimization must be continued and in the case of the former, we correct the eigenfrequency from the obtained perturbations by determining the curvature from fitting. This is based on the harmonic approximation of the potential energy surface $E_{harm}$ at the point $x_0$, that is either an obtained transition state or a minimum. $\overleftrightarrow{H}$ is the hessian and $\overleftrightarrow{M}$ the reduced mass matrix. For small deviations along an eigenmode j $(x_0 + a\vec{u}_j)$, this reduces to:

$$E_{harm}(x_0 + a\vec{u}_j) = E(x_0) + \frac{1}{2}a^2 \vec{u}_j^T \overleftrightarrow{H} \vec{u}_j \quad \text{(Equation 2)}$$

$$E_{harm}(x_0 + a\vec{u}_j) = E(x_0) + \frac{1}{2}a^2 \vec{u}_j^T \overleftrightarrow{M}^{-1}(\overleftrightarrow{M}\overleftrightarrow{H}) \vec{u}_j \quad \text{(Equation 3)}$$

$$E_{harm}(x_0 + a\vec{u}_j) = E(x_0) + \frac{1}{2}a^2 \vec{u}_j^T \overleftrightarrow{M}^{-1} \omega_j^2 \vec{u}_j \quad \text{(Equation 4)}$$

$$E_{harm}(a) = E(x_0) + \frac{1}{2}c_j a^2 \quad \text{with} \quad c_j = \omega_j^2 m_j \quad \text{and} \quad m_j = \vec{u}_j^T \overleftrightarrow{M}^{-1} \vec{u}_j \quad \text{(Equation 5)}$$

This now enables translating the curvature $c_{j,fit}$ of the perturbed data (force constant) into a vibration frequency via $\omega_{j,fit} = \sqrt{c_{j,fit}/m_j} = 2\pi \nu_{j,fit}$.

Figure S10 visualizes the results of this approach for all instable modes of minima and transition states. Three of the calculated transition states exhibit an additional instable mode. Also, for three of the minima one instable frequency was found. But when comparing the curvatures from the vibrational analyses with the distinct perturbations along the instable modes, we identify all additional instable vibration frequencies to result from numerical inaccuracies. By harmonically fitting the energies of the perturbed points all obtained vibration frequencies were corrected.

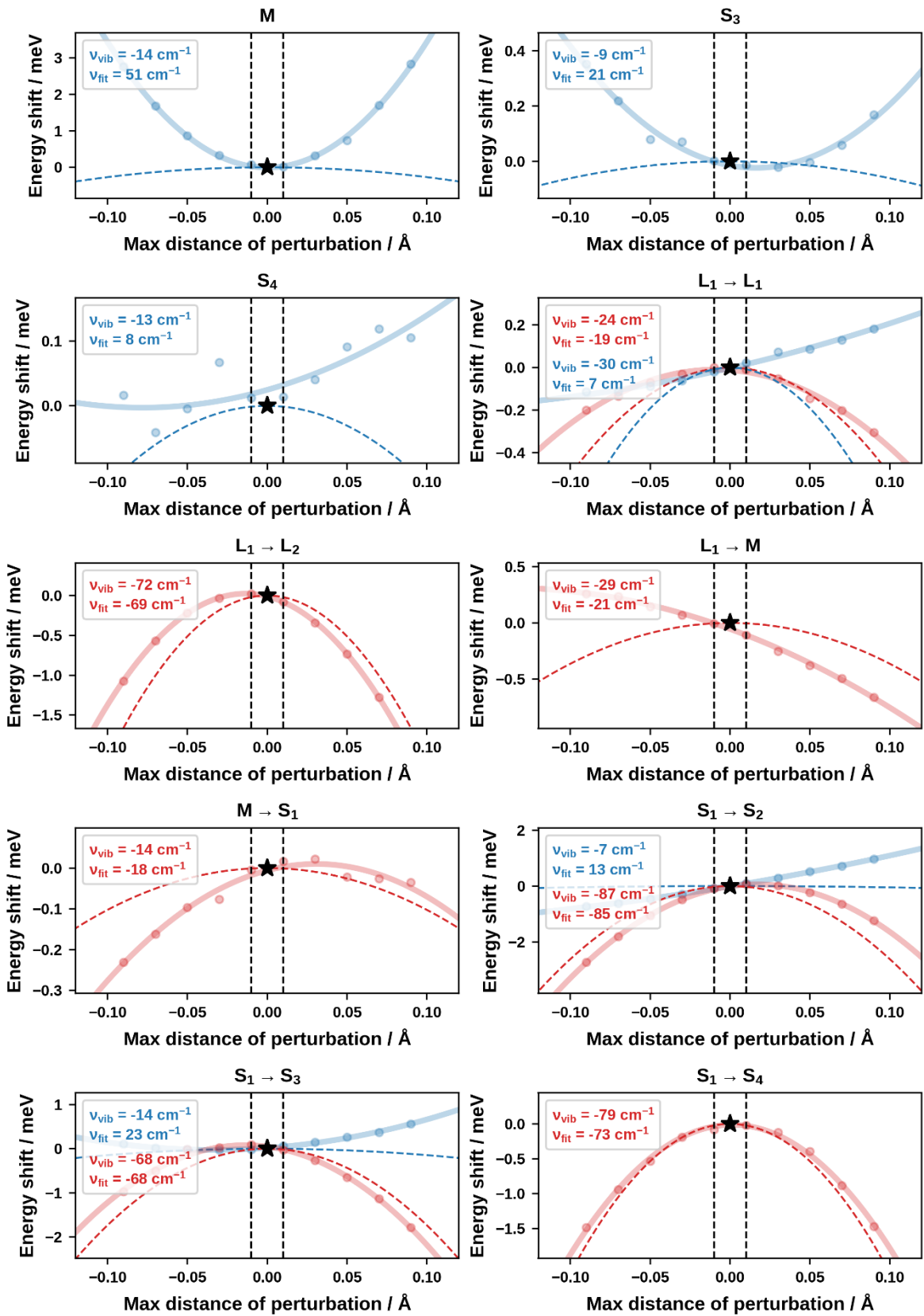

**Figure S10:** *Displacements along all instable vibration modes appearing at minima and states. A schematic description of the content is provided in Figure S9.*

### 4.2.3 Uncertainty estimation of vibrational frequencies

The uncertainty estimation of the computed vibrational frequencies depends also the quality of the obtained extremum. For the case that the computed extremum lies within the harmonic range of the real extremum, the curvatures and vibration frequencies should be constant. In practice, numerical instabilities, which are caused by DFT, can additionally distort vibration frequencies (for details see Section 4.2.2).

To get a rough estimate of the uncertainty of vibrational frequencies, the process described in the previous section was repeated for six eigenmodes with frequencies from 11 to 2218 cm$^{-1}$, as displayed in Figure S11. This was done for the rate-limiting transition state of the joint process of standing up (L$_1$→S$_1$).

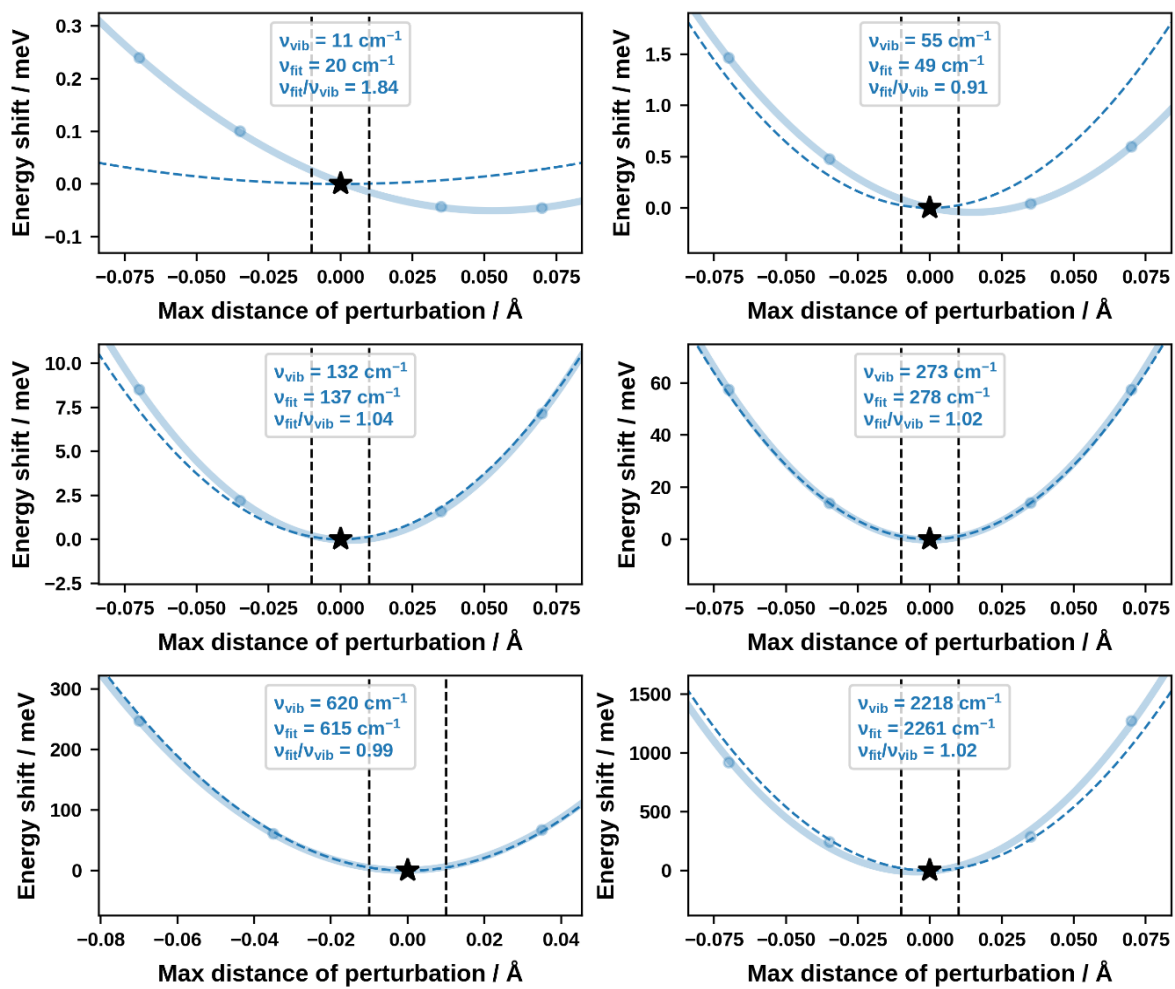

**Figure S11:** *Perturbations along different vibration modes of the rate-limiting transition state of L$_1$→S$_1$. A schematic description of the content is provided in Figure S9.*

## 4.3 Attempt frequencies

The attempt frequency $A$ is obtained via the stable frequencies of the initial ($v_i^{ini}$) and the transition state ($v_i^{TS}$), as stated in Equation 6. The ansatz and the result of the relative uncertainty are given by Equation 7 and 8, respectively.

$$A = \frac{\prod_{i=1}^{3N} v_i^{ini}}{\prod_{i=1}^{3N-1} v_i^{TS}} = \frac{P^{ini}}{P^{TS}} \qquad \text{with } P = \prod_i v_i \qquad \text{(Equation 6)}$$

$$\Delta A = \sqrt{\left(\frac{\partial A}{\partial P^{ini}}\right)^2 \cdot (\Delta P^{ini})^2 + \left(\frac{\partial A}{\partial P^{TS}}\right)^2 \cdot (\Delta P^{TS})^2} \qquad \text{with } \frac{\Delta P}{P} = \sqrt{\sum_i \left(\frac{\Delta v_i}{v_i}\right)^2} \qquad \text{(Equation 7)}$$

$$\frac{\Delta A}{A} = \sqrt{\sum_{i=1}^{3N} \left(\frac{\Delta v_i^{ini}}{v_i^{ini}}\right)^2 + \sum_{i=1}^{3N-1} \left(\frac{\Delta v_i^{TS}}{v_i^{TS}}\right)^2} \qquad \text{(Equation 8)}$$

The relative uncertainty of the attempt frequency solely depends on the relative uncertainties of the computed vibration frequencies. Based on the results of Figure S11 and the influence of conducting the vibrational analysis for the results of the 4x4 super cell without substrate relaxation, we estimate the relative uncertainty of the attempt frequencies to $\frac{\Delta A}{A} \approx 10$.

## 4.4 Suppression temperatures

In the main paper temperatures $T_{supp}$ are proposed that sufficiently suppress transitions of individual molecules. In detail, the temperatures $T_{supp}$ are determined for transition rates of $k_{supp} = 1$ s$^{-1}$ via Equation 9 (energy barrier $\Delta E^{\ddagger}$, the attempt frequency $A$).

$$T_{supp} = \frac{\Delta E^{\ddagger}}{k_B (\ln(A) - \ln(k_{supp}))} \qquad \text{(Equation 9)}$$

To get a rough feeling for the uncertainties of the suppression temperatures we perform a simple error propagation. The general approach and the resulting description are shown in Equations 10 and 11, respectively.

$$\Delta T_{supp} = \sqrt{\left(\frac{\partial T_{supp}}{\partial \Delta E^{\ddagger}}\right)^2 \cdot (\Delta \Delta E^{\ddagger})^2 + \left(\frac{\partial T_{supp}}{\partial A}\right)^2 \cdot (\Delta A)^2 + \left(\frac{\partial T_{supp}}{\partial k_{supp}}\right)^2 \cdot (\Delta k_{supp})^2} \qquad \text{(Equation 10)}$$

$$\Delta T_{supp} = T_{supp} \cdot \sqrt{\left(\frac{\Delta \Delta E^{\ddagger}}{\Delta E^{\ddagger}}\right)^2 + \left(\frac{1}{\ln(A) - \ln(k_{supp})} \cdot \frac{\Delta A}{A}\right)^2 + \left(\frac{1}{\ln(A) - \ln(k_{supp})} \cdot \frac{\Delta k_{supp}}{k_{supp}}\right)^2} \qquad \text{(Equation 11)}$$

In Equation 11, it is evident that the uncertainty of the suppression temperatures grows with its nominal value. For all transitions $\Delta \Delta E^{\ddagger} = 0.04$ eV is assumed, which corresponds approximately to the double of the uncertainty of the adsorption energies obtained via DFT, and a relative error of one order of magnitude for A, i.e. $\frac{\Delta A}{A} = 10$ (for more details see Section 4.3). $\Delta k_{supp}$ is not accounted for ($\Delta k_{supp} = 0$).

The results for the forward and reverse transitions are visualized in Figure S12 and stated in Table S5 and S6, respectively. Therein, the nominal values ($T_{supp}$) and the uncertainty estimate ($\Delta T_{supp}$) of the suppression temperature are stated, as well as the single uncertainty contributions regarding the attempt frequency and the energy barrier.

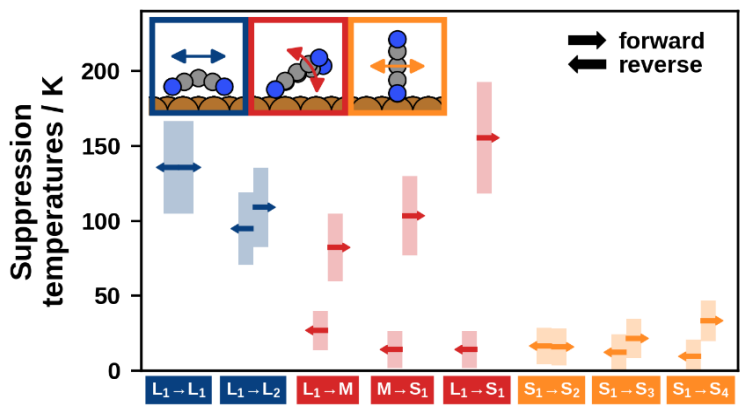

**Figure S12:** *Estimated temperatures for suppression obtained for transition rates of 1 day$^{-1}$. The arrow directions indicate the suppression temperatures required for forward or reverse transitions. Shaded areas indicate the temperature uncertainties connected to the respective transition.*

**Table S5:** *Uncertainties of suppression temperatures of forward transitions*

| Transition | $T_{\text{supp}}$ / K | $\Delta T_{\text{supp}}$ / K | $\left|\frac{\partial T_{\text{supp}}}{\partial A} \cdot \Delta A\right|$ / K | $\left|\frac{\partial T_{\text{supp}}}{\partial \Delta E^{\ddagger}} \cdot \Delta \Delta E^{\ddagger}\right|$ / K |
|---|---|---|---|---|
| $L_1 \rightarrow L_1$ | 136 | 31 | 29 | 10 |
| $L_1 \rightarrow L_2$ | 109 | 27 | 24 | 10 |
| $L_1 \rightarrow M$ | 82 | 23 | 20 | 11 |
| $M \rightarrow S_1$ | 103 | 26 | 24 | 11 |
| $L_1 \rightarrow S_1$ | 155 | 37 | 36 | 11 |
| $S_1 \rightarrow S_2$ | 16 | 12 | 4 | 12 |
| $S_1 \rightarrow S_3$ | 22 | 13 | 5 | 12 |
| $S_1 \rightarrow S_4$ | 33 | 14 | 8 | 11 |

**Table S6:** *Uncertainties of suppression temperatures of reverse transitions*

| Transition | $T_{\text{supp}}$ / K | $\Delta T_{\text{supp}}$ / K | $\left|\frac{\partial T_{\text{supp}}}{\partial A} \cdot \Delta A\right|$ / K | $\left|\frac{\partial T_{\text{supp}}}{\partial \Delta E^{\ddagger}} \cdot \Delta \Delta E^{\ddagger}\right|$ / K |
|---|---|---|---|---|
| $L_1 \rightarrow L_1$ | 136 | 31 | 29 | 10 |
| $L_2 \rightarrow L_1$ | 95 | 24 | 22 | 11 |
| $M \rightarrow L_1$ | 27 | 13 | 7 | 11 |
| $S_1 \rightarrow M$ | 14 | 12 | 4 | 12 |
| $S_1 \rightarrow L_1$ | 14 | 12 | 4 | 12 |
| $S_2 \rightarrow S_1$ | 17 | 12 | 4 | 11 |
| $S_3 \rightarrow S_1$ | 12 | 12 | 3 | 12 |
| $S_4 \rightarrow S_1$ | 10 | 11 | 2 | 11 |